\documentclass{article}
\usepackage[T1]{fontenc} % add special characters (e.g., umlaute)
\usepackage[utf8]{inputenc} % set utf-8 as default input encoding
\usepackage{ismir,amsmath,cite,url}
\usepackage{graphicx}
\usepackage{color}

\usepackage{amsfonts}
\usepackage{makecell}
\usepackage{subcaption}
\usepackage{multirow}
\usepackage{cleveref}
\usepackage{textgreek}

\usepackage{lineno}
\title{Sketching the Expression: \\
Flexible Rendering of Expressive Piano Performance with Self-Supervised Learning}

\multauthor
{Seungyeon Rhyu$^1$ \hspace{1cm} Sarah Kim$^2$ \hspace{1cm} Kyogu Lee$^{1,3}$} {
$^1$ Music and Audio Research Group (MARG), Seoul National University, South Korea\\
$^2$ Krust Universe, South Korea\\
$^3$ Graduate School of AI, AI Institute, Seoul National University, South Korea\\
{\tt\small rsy1026@snu.ac.kr, estelle.kim@krustuniverse.com, kglee@snu.ac.kr}
}

\def\authorname{S. Rhyu, S. Kim, and K. Lee}

\begin{document}

\maketitle
\begin{abstract}
We propose a system for rendering a symbolic piano performance with flexible musical expression. It is necessary to actively control musical expression for creating a new music performance that conveys various emotions or nuances. However, previous approaches were limited to following the composer's guidelines of musical expression or dealing with only a part of the musical attributes. We aim to disentangle the entire musical expression and structural attribute of piano performance using a conditional VAE framework. It stochastically generates expressive parameters from latent representations and given note structures. In addition, we employ self-supervised approaches that force the latent variables to represent target attributes. Finally, we leverage a two-step encoder and decoder that learn hierarchical dependency to enhance the naturalness of the output. Experimental results show that our system can stably generate performance parameters relevant to the given musical scores, learn disentangled representations, and control musical attributes independently of each other.
\end{abstract}

\section{Introduction}\label{sec:introduction}

Computational modeling of expressive music performance focuses on mimicking human behaviors that convey the music \cite{widmer2004computational,chacon2018computational}. For piano performance, one common task is to \textit{render} an expressive performance from a quantized musical score. It aims to reproduce the loudness and timing of musical notes that fits to the given score. Most of the conventional studies have used musical scores of Western piano music that includes sufficient amount of guidelines for musical expressions \cite{widmer2009yqx,kim2013statistical,chacon2015evaluation,chacon2017evaluation}. Recent studies using deep learning methods have successfully rendered plausible piano performances that are comparable to those of professional pianists from the given Classical scores \cite{maezawa2018deep,jeong2019virtuosonet,jeong2019graph}.

More recently, it has increased attention to \textit{controlling} music performance by manipulating one or more \textit{disentangled} representations from a generative model. These representations are sensitive to the variation of certain factors while invariant to other factors \cite{bengio2013representation}. Maezawa \emph{et al.} aimed to control a performer's interpretation through a conditional variational recurrent neural network (CVRNN) \cite{maezawa2019rendering}. They intended to disentangle a time-variant representation of the personal interpretation. In the acoustic domain, Tan \emph{et al.} proposed a generative model based on a Gaussian mixture variational autoencoder (GM-VAE) that separately controlled dynamics and articulations of the notes \cite{tan2020generating}. Their novelty lied in learning multiple representations of high-level attributes from the low-level spectrogram.

However, these studies have constrained musical creativity. Maezawa \emph{et al.} controlled musical expression only through quantized features from the musical scores. Tan \emph{et al.} did not consider controlling tempo or timing with a latent representation. These methods may have restricted any potential for rendering piano performances with flexible musical expression. Musical creativity can be expanded not only by composers but also by performers who can elastically choose various strategies to highlight multiple nuances or emotions \cite{bresin2000emotional,livingstone2010changing,bernays2014investigating}. Moreover, the music generation field can be also broadened if static music created by automatic composition systems can be easily colored with realistic and elastic expression \cite{oore2020thistime}.

Therefore, we attempt a new approach that renders piano performances with flexible musical expressions. We disregard a typical assumption from previous studies that a performer must follow a composer's intent  \cite{bhatara2011perception,friberg2006overview,flossman2013expressive,kim2013statistical}. According to the literature, performers learn to identify or imitate "expressive models", or \emph{explicit planning}, of existing piano performances \cite{woody1999relationship}. We focus on this attribute, defining it as a higher-level \emph{sketch} of the expressive attributes (i.e. dynamics, articulation, and tempo \cite{lerch2019music}) that the performer draws based on a personal interpretation of the musical piece \cite{woody1999relationship,kim2013statistical,maezawa2019rendering}. We also assume that the remaining attribute represents common performing strategies that are connected to certain musical patterns, while these strategies slightly differ across performers \cite{repp1999microcosm,honing2001from}. We call this attribute as a \emph{structural attribute} that belongs to given note structures of a musical piece.

In this study, we propose a generative model that can flexibly control the entire musical expression, or the explicit planning, of symbolic piano performance\footnote{\url{https://github.com/rsy1026/sketching_piano_expression}}. Our system is based on a conditional variational autoencoder (CVAE) that is modified for sequential data \cite{zhu2020s3vae,maezawa2019rendering}. The system generates multiple parameters of piano performance from a note structure of a musical passage, using disentangled representations for the explicit planning and structural attribute.

We employ a self-supervised learning framework to force the latent representations to learn our target attributes \cite{locatello2019challenging,hendrycks2019using,zhu2020s3vae}.
In addition, we facilitate independent control of the three expressive attributes--dynamics, articulation, and tempo--by utilizing an existing method that aligns the latent code with a target attribute \cite{pati2020attribute,tan2020music}. Finally, we design a novel mechanism that intuitively models a polyphonic structure of piano performance. In particular, we insert intermediate steps for \emph{chordwise} encoding and decoding of the piano performance to our encoder-decoder architecture, where a \emph{chord} denotes a group of simultaneous notes.

Our approach has several contributions as follows: 
1) Our system aims to control musical expression while maintaining any characteristics induced by a given musical structure;
2) We use self-supervised learning where new supervisory signals are involved in regularizing the latent representations effectively;
3) Our system aims to control multiple expressive attributes independently of each other;
4) Lastly, we leverage an intermediate step that projects a notewise representation into the chordwise in the middle of our system to intuitively model the polyphonic structure of piano performance.

\section{Proposed Methods}

We aim to build a generative model that factorizes expressive piano performance as the explicit planning and structural attribute. The model is based on a conditional variational autoencoder (CVAE) that reproduces performance parameters based on a given musical structure. 

% In this section, we describe our proposed methods as follows: First, we introduce data representations for a piano performance and its musical score. Second, we introduce a mechanism for modeling the musical hierarchy. Third, we describe the overall architecture of our proposed system. We also introduce several auxiliary tasks for enhancing the disentanglement of the target representations. Lastly, we summarize the entire training objective.

\subsection{Data Representation}

We extract features that represent a human performance and the corresponding musical score, following the conventional studies \cite{flossman2013expressive,jeong2019score,maezawa2019rendering}. 
% A musical score is represented as the quantized temporal, vertical positions of notes within the score. A human performance is represented as the three expressive attributes, dynamics, articulations, and tempo. The ranges of all feature values are preliminarily determined from the collected data.

\textbf{Performance Features.} We extract three features that represent the expressive attributes of each performed note, respectively: 
\textbf{MIDIVelocity} is a MIDI velocity value that ranges from 24 to 104.
\textbf{IOIRatio} represents an instantaneous variation in tempo. We compute an inter-onset-interval (IOI) between the onset of a note and the mean onset of the \textit{previous} chord for both a performed note and the corresponding score note. Then, a ratio of performed IOI to score IOI is calculated, clipped between 0.125 and 8, and converted into a logarithmic scale \cite{kim2013statistical}.  
\textbf{Articulation} represents how much a note is shortened or lengthened compared to the instantaneous tempo. It is a ratio of a performed duration to an IOI value between the onset of a note and mean onset of the \textit{next} chord \cite{flossman2013expressive}. It is clipped between 0.25 and 4 and converted into a logarithmic scale.

\textbf{Score Features.} The features for a musical score represent eight categorical attributes for how the notes are composed:
\textbf{Pitch} is a MIDI index number that ranges from 21 to 108. 
\textbf{RelDuration} and \textbf{RelIOI} are 11-class attributes of a quantized duration and IOI between a note onset and a previous chord, respectively. They range from 1 to 11, and each class represents a multiple of a 16th note's length with respect to a given tempo \cite{roberts2017hierarchical,dong2018musegan}. 
\textbf{IsTopVoice} is a binary attribute of whether the note is the uppermost voice. It is heuristically computed regarding pitches and durations of surrounding notes.
\textbf{PositionInChord} and \textbf{NumInChord} are 11-class attributes of a positional index of a note within its chord and the total number of notes in that chord, respectively, that range from 1 to 11. An index 1 for PositionInChord denotes the most bottom position.
\textbf{Staff} is a binary attribute of the staff of a note, either of the G clef or F clef.
\textbf{IsDownbeat} is a binary attribute of whether a note is at a downbeat or not.

\subsection{Modeling Musical Hierarchy}

Inspired by previous studies \cite{kim2013statistical,jeong2019graph,jeong2019virtuosonet,wu2021musemorphose}, we build a two-step encoder and decoder: An encoder models both notewise and chordwise dependencies of the inputs, and a decoder reconstructs the notewise dependency from the chordwise representation and the notewise condition. We denote a \textit{chord} as a group of notes that are hit simultaneously, regardless of the staff, so that they sound together at an instant time \cite{wang2020pianotree}. Thus, learning the chordwise dependency is analogous to direct modeling of the temporal progression of the piano performance. Let $\mathcal{M} \in \mathbb{R}^{C\times N}$ be a matrix that aligns serialized notes to their polyphonic structure, where $C$ and $N$ are the number of chords and the number of notes, respectively. Within the encoder, the notewise representation is sequentially average-pooled by $\mathcal{M}$ with dynamic kernel sizes where each size represents the number of notes in each chord. We denote this operation as \textit{N2C}. In this way, we can directly model chord-level dependency of the note-level expressive parameters \cite{wu2021musemorphose}. In contrast, the decoder extends the chordwise representation from the encoder back to the notewise using the transposed alignment matrix $\mathcal{M}^T$, of which process we denote as \textit{C2N}. Along this, the notewise embedding of the score features replenishes the notewise information for the output. Consequently, notes in the same chord \emph{share} any information of their corresponding chord, while maintaining their differences by the conditional score features: 
\begin{equation}
    \text{N2C}(e) = \frac{\mathcal{M}\cdot e}{\textstyle\sum_{n=1}^{N}\mathcal{M}_{n,1:C}},
    \quad
    \text{C2N}(e) = \mathcal{M}^{\text{T}} \cdot e 
\label{eq:c2n}
\end{equation}
where $e$ denotes a notewise or chordwise representation.

\subsection{Overall Network Architecture}

\begin{figure} % [hbt!]
\centering
\includegraphics[width=0.99\columnwidth]{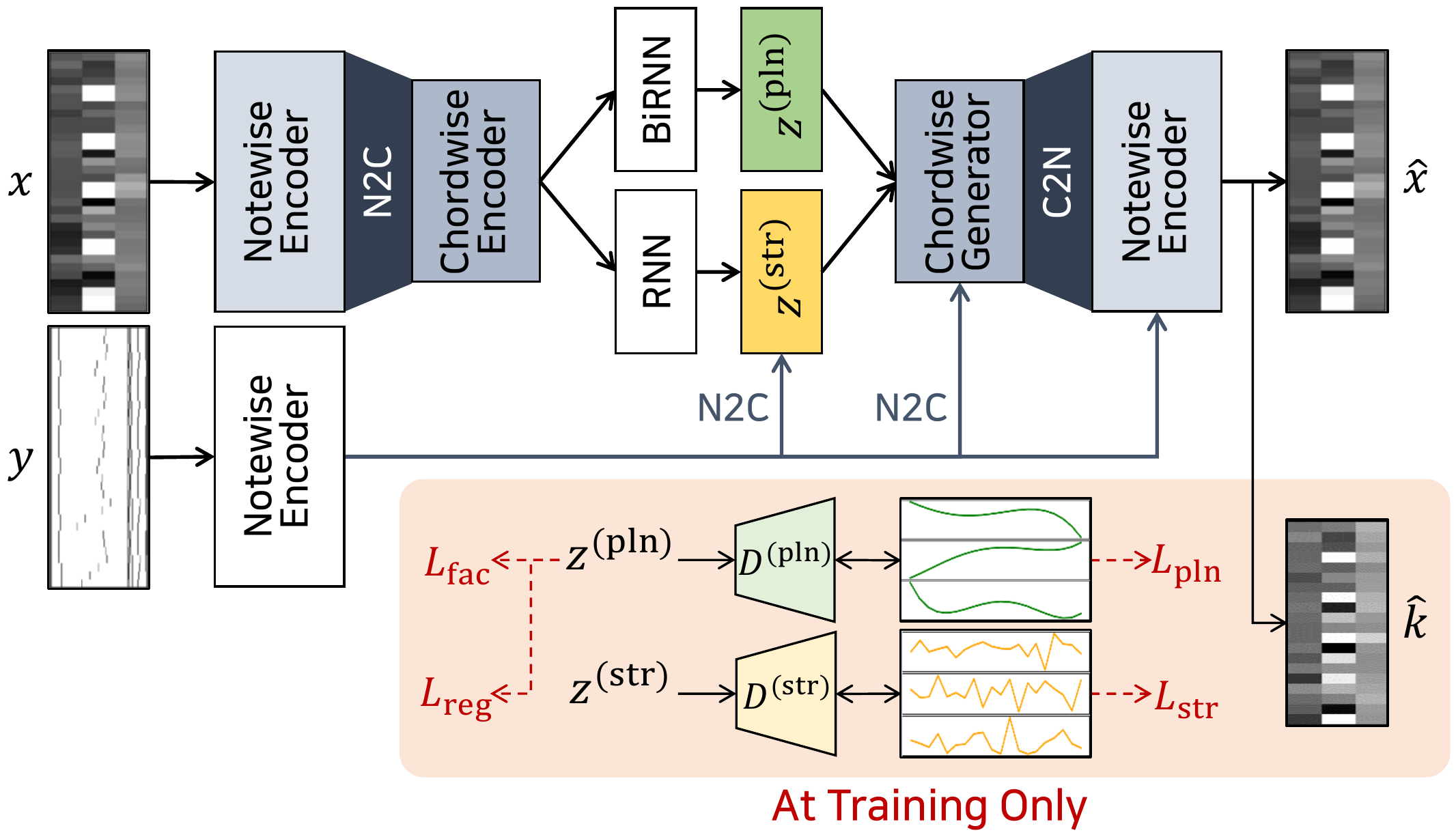}
\caption{Overall architecture of the proposed system. The orange box includes the auxiliary tasks only for training.}
\label{fig:model}
\end{figure}

Our proposed network is generally based on the conditional VAE framework \cite{kingma2013auto,sohn2015learning}. Concretely, we use the sequential VAE that is modified for generation of sequential data \cite{li2018disentangled,zhu2020s3vae,maezawa2019rendering}. Let $x=\{x_{n}\}_{n=1}^N$ be a sequence of the performance features, and $y=\{y_{n}\}_{n=1}^N$ be a sequence of the conditional score features. Our network has two \emph{chordwise} latent variables $z^{(\text{pln})}=\{z^{(\text{pln})}_{c}\}_{c=1}^C\in \mathbb{R}^{C\times d^{(\text{pln})}}$ and $z^{(\text{str})}=\{z^{(\text{str})}_{c}\}_{c=1}^C\in \mathbb{R}^{C\times d^{(\text{str})}}$ that represent explicit planning and structural attribute, where $d^{(\text{pln})}$ and $d^{(\text{str})}$ are the sizes of $z^{(\text{pln})}$ and $z^{(\text{str})}$, respectively. Our network generates notewise performance parameters $x$ from these latent variables and given score features $y$. 
The overall architecture of our proposed system is illustrated in \figref{fig:model}.

{\bf Generation.} A probabilistic generator parameterized by $\theta$ produces the note-level performance parameters $x$ from the two latent variables $z^{(\text{pln})}$ and $z^{(\text{str})}$ with the given condition $y$. We note that the latent variables are in chord-level. This decreases a computational cost and also enables intuitive modeling of polyphonic piano performance where each time step represents a stack of notes and the simultaneous notes share common characteristics \cite{jeong2019virtuosonet}:
\begin{equation}
\begin{split}
    p_{\theta}(x,y,z^{(\text{pln})},z^{(\text{str})})=&p_{\theta}(x|z^{(\text{pln})},z^{(\text{str})},y) \\
    p_{\theta}(z^{(\text{pln})})&\prod_{c=1}^C p_{\theta}(z^{(\text{str})}_{c}|z^{(\text{str})}_{<c},y^{(\text{chd})}_{\leq c})
\end{split}
\end{equation}
where $y^{(\text{chd})} = \text{N2C}(e_{y})$ is the chordwise embedding, and $e_{y}$ is the notewise embedding for $y$. We assume that the prior of $z^{(\text{pln})}_{c}$ is a standard normal distribution. In contrast, $z^{(\text{str})}_{c}$ is sampled from a sequential prior \cite{chung2015recurrent,li2018disentangled,zhu2020s3vae}, conditioned on both previous latent variables and chordwise score features: $z^{(\text{str})}_{c}\sim \mathcal{N}(\mu^{(\text{prior})},\text{diag}(\sigma^{(\text{prior})^2})$, where $[\mu^{(\text{prior})},\sigma^{(\text{prior})}]=f^{(\text{prior})}(z^{(\text{str})}_{<c},y^{(\text{chd})}_{\leq c})$, and $f^{(\text{prior})}$ is a unidirectional recurrent neural network. 
The latent representations and $y^{(\text{chd})}$ pass through the decoder as shown in \figref{fig:model}. During training, the model predicts the intermediate chordwise output that is computed as $\text{N2C}(x)$. This is to enhance reconstruction power of our system, propagating accurate information of chord-level attributes to the final decoder. The intermediate activation is then extended to the notewise through the C2N operation. The note-level parameters are generated autoregressively based on this activation and the notewise score feature. We use teacher forcing during training \cite{williams1989learning}.

{\bf Inference.} A probabilistic encoder parameterized by $\phi$ approximates the posterior distibutions of the latent representations $z^{(\text{pln})}$ and $z^{(\text{str})}$ from the performance input $x$ and conditional score input $y$:
\begin{equation}
\begin{aligned}
    q_{\phi}(z^{(\text{pln})},z^{(\text{str})}|x,y)=&q_{\phi}(z^{(\text{pln})}|x^{(\text{chd})})\\
    &\prod_{c=1}^C q_{\phi}(z^{(\text{str})}_{c}|x^{(\text{chd})}_{\leq c},y^{(\text{chd})}_{\leq c})
\end{aligned}
\end{equation}
where $x^{(\text{chd})} = \text{N2C}(e_{x})$ is the chordwise embedding, and $e_{x}$ is the notewise embedding for $x$. The posterior distributions of $z^{(\text{pln})}_{c}$ and $z^{(\text{str})}_{c}$ are approximated by distribution parameters encoded by $f^{(\text{pln})}(x^{(\text{chd})})$ and $f^{(\text{str})}(x^{(\text{chd})}, y^{(\text{chd})})$, where $f^{(\text{pln})}$ and $f^{(\text{str})}$ are bidirectional and unidirectional recurrent neural networks, respectively. 
% \begin{equation}
%     z^{(\text{pln})}_{c} \sim \mathcal{N}(\mu^{(\text{pln})}, \text{diag}({\sigma^{2}}^{(\text{pln})})),
%     \quad
%     z^{(\text{str})}_{c} \sim \mathcal{N}(\mu^{(\text{str})}, \text{diag}({\sigma^{2}}^{(\text{str})}))
% \end{equation}
% where $[\mu^{(\text{pln})}, \sigma^{(\text{pln})}]=f^{(\text{pln})}(x^{(\text{chd})})$, $ [\mu^{(\text{str})}, \sigma^{(\text{str})}]=f^{(\text{str})}(x^{(\text{chd})}, y^{(\text{chd})})$, and $f^{(\text{pln})}$ and $f^{(\text{str})}$ are bidirectional and unidirectional recurrent neural networks, respectively. 
We note that $z^{(\text{pln})}$ is independent of the score features $y$. This allows a flexible transfer of the explicit planning among other musical pieces. On the other hand, $z^{(\text{str})}$ is constrained by $y$ since the structural attributes are dependent on the note structure. 

{\bf Training.} We train the models $p_{\theta}$ and $q_{\phi}$ by approximating marginal distributions of the performance features $x$ conditioned on the score features $y$. This requires to maximize negative evidence lower bound (ELBO) that includes regularization force by Kullback–Leibler divergence \cite{kingma2013auto}:
\begin{equation}
\begin{aligned}
&\mathcal{L}_{\text{VAE}}= \mathbb{E}_{q_{\phi}(z^{(\text{pln})},z^{(\text{str})}|x,y)}\left[\log p_{\theta}(x|z^{(\text{pln})},z^{(\text{str})},y)\right]\\ 
&+ \mathbb{E}_{q_{\phi}(z^{(\text{pln})},z^{(\text{str})}|x,y)}\left[\log p_{\theta}(k|z^{(\text{pln})},z^{(\text{str})},y)\right]\\
&-\text{KL}(q_{\phi}(z^{(\text{pln})}|x)\Vert p_{\theta}(z^{(\text{pln})})) \\
&-\sum_{c=1}^C\text{KL}(q_{\phi}(z^{(\text{str})}_{c}|x^{(\text{chd})}_{\leq c},y^{(\text{chd})}_{\leq c})\Vert p_{\theta}(z^{(\text{str})}_{c}|z^{(\text{str})}_{<c},y^{(\text{chd})}_{\leq c}))
\label{eq:vae_loss} 
\end{aligned}
\end{equation}
where $k=\text{N2C}(x)$ is the chordwise performance features.

\subsection{Regularizing the Latent Variables}

We enhance disentanglement of the latent representations $z^{(\text{pln})}$ and $z^{(\text{str})}$ using four regularization tasks \cite{zhu2020s3vae}. 

% First of all, we employ two prediction tasks where the two latent variables are directly involved. We also regularize the latent variables to ensure that they do not interfere with each other. Finally, we constrain certain dimensions of $z^{(\text{pln})}$ for independent control of the expressive attributes.

{\bf Prediction Tasks.} We extract new supervisory signals for additional prediction tasks from the input data \cite{zhu2020s3vae}. We define a signal of explicit planning $I^{(\text{pln})}$ as a set of smoothed contours of the expressive parameters. It is extracted as a polynomial function predicted from the chordwise performance parameters $k$. We also derive a signal of structural attribute as $I^{(\text{str})}=\text{sign}(k-I^{(\text{pln})})$ which represents normalized directions of the performance parameters.
We train two discriminators $D^{(\text{pln})}$ and $D^{(\text{str})}$ that directly receive $z^{(\text{pln})}$ and $z^{(\text{str})}$, respectively. $D^{(\text{pln})}$ is composed of $A$ sub-discriminators where each discriminator $D^{(\text{pln})}_{a}$ predicts a signal $I^{(\text{pln})}_{a}$ for each expressive attribute $a$ from $z^{(\text{pln})}_{a}\in \mathbb{R}^{C\times (d^{(\text{pln})}/A)}$, where $z^{(\text{pln})}_{a}$ is a constituent part of $z^{(\text{pln})}$, and $A$ is the number of expressive attributes. This setting is for a clear disentanglement among the expressive attributes. On the other hand, $D^{(\text{str})}$ predicts the signal $I^{(\text{str})}$ at once for all expressive attributes that belong to the same musical structure. All discriminators are jointly trained with the generative model, and the costs $\mathcal{L}_{\text{pln}}$ and $\mathcal{L}_{\text{str}}$ are minimized as $\mathcal{L}_{\text{pln}}= \frac{1}{A}\sum_{a}\text{MSE}(D^{(\text{pln})}_{a}(z^{(\text{pln})}_{a}), I^{(\text{pln})}_{a})$ and $\mathcal{L}_{\text{str}}=\text{MSE}(D^{(\text{str})}(z^{(\text{str})}), I^{(\text{str})})$, respectively.
% \begin{equation}
% \begin{split}
%     \mathcal{L}_{\text{pln}} &= \frac{1}{A}\sum_{a}\text{MSE}(D^{(\text{pln})}_{a}(z^{(\text{pln})}_{a}), I^{(\text{pln})}_{a})
% \label{eq:loss_pln} 
% \end{split}
% \end{equation}
% \begin{equation}
% \begin{split}
%     \mathcal{L}^{(\text{str})} &= \text{MSE}(D^{(\text{str})}(z^{(\text{str})}), I^{(\text{str})}))
% \label{eq:loss_str} 
% \end{split}
% \end{equation}

{\bf Factorizing Latent Variables.} We further constrain a generator to guarantee that $z^{(\text{pln})}$ delivers correct information regardless of $z^{(\text{str})}$ \cite{hu2017toward}. During training, we sample a new output $\tilde{x}$ using $z^{(\text{pln})} \sim q_\phi(z^{(\text{pln})}|x)$ and $\tilde{z}^{(\text{str})} \sim p_\theta(z^{(\text{str})})$. Then, we re-infer $\tilde{z}^{(\text{pln})} \sim q_\phi(\tilde{z}^{(\text{pln})}|\tilde{x})$ to estimate the superversory signal $I^{(\text{pln})}$. This prediction loss is backpropagated only through the generator:   
\begin{equation}
    \mathcal{L}_{\text{fac}} = \frac{1}{A}\sum_{a}\text{MSE}(D^{(\text{pln})}_{a}(\tilde{z}^{(\text{pln})}_{a}), I^{(\text{pln})}_{a})
\label{eq:fac} 
\end{equation}

{\bf Aligning Latent Variables with Factors.} Finally, we enable the "sliding-fader" control of the expressive attributes \cite{tan2020music}. To this end, we employ the regularization loss proposed by Pati \emph{et al.} \cite{pati2020attribute} that aligns specific dimensions of $z^{(\text{pln})}$ with the target expressive attributes. This method assumes that a latent representation can be disentangled through its monotonic relationship with a target attribute. Let $d_{i}$ and $d_{j}$ be a target dimension $d$ of $i$th and $j$th latent representations, respectively, where $d\in z^{(\text{pln})}_{a}$, $i,j\in [1,M]$, and $M$ is the size of a mini-batch. A distance matrix $\mathcal{D}_{d}$ is computed between $d_{i}$ and $d_{j}$ within a mini-batch, where $\mathcal{D}_{d}=d_{i} - d_{j}$. A similar distance matrix $\mathcal{D}_{a}$ is computed for the two target attribute values $a_{i}$ and $a_{j}$. We minimize a MSE between $\mathcal{D}_{d}$ and $\mathcal{D}_{a}$ as follows: 
\begin{equation}
    \mathcal{L}_{\text{reg}} 
    = \text{MSE}(\text{tanh}(\mathcal{D}_{d}),\text{sign}(\mathcal{D}_{a}))
\label{eq:reg_c} 
\end{equation}

\begin{figure*}
\centering
\begin{subfigure}{0.33\textwidth}
    \includegraphics[width=\textwidth]{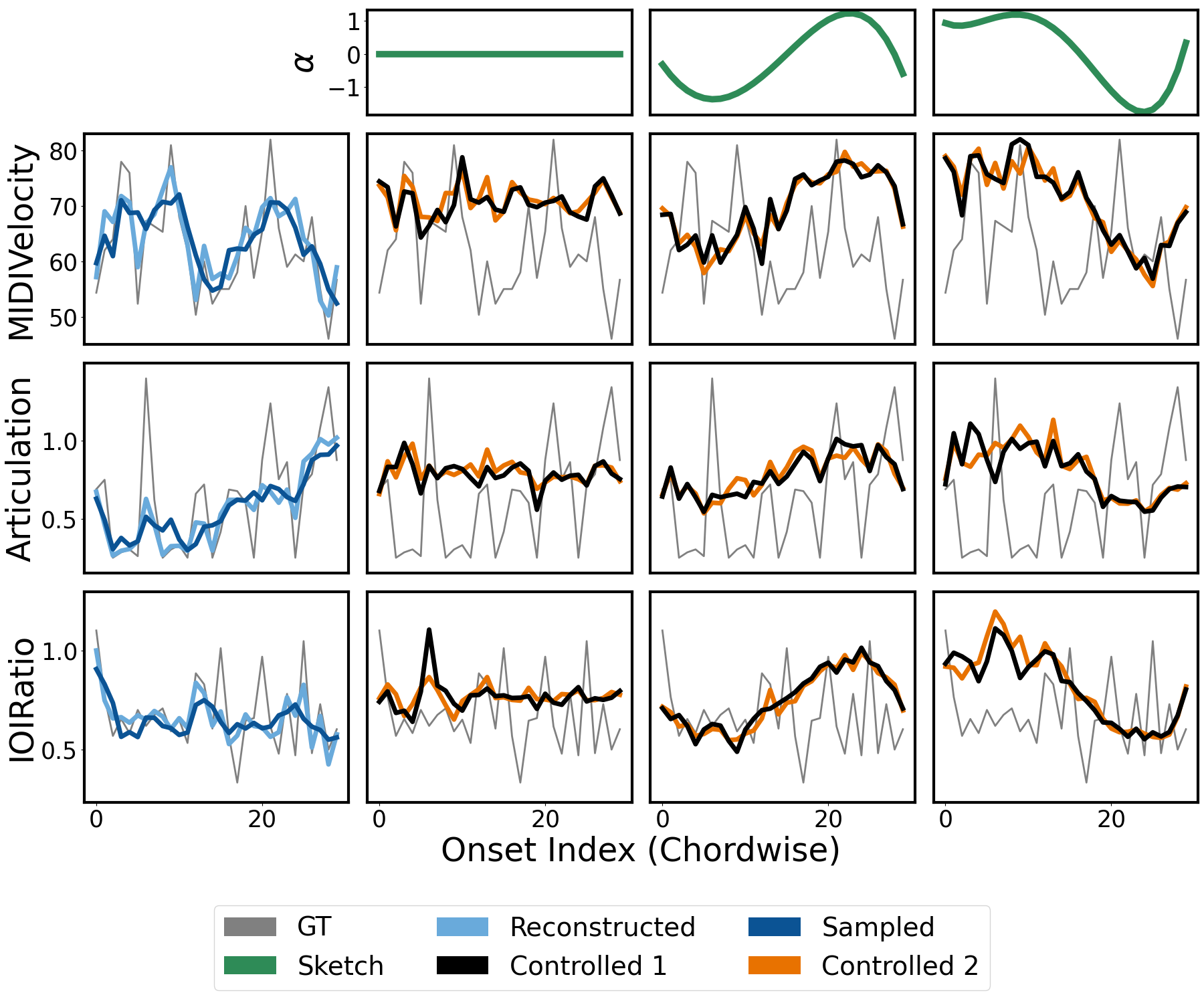}
    \caption{}
    \label{fig:sketch1}
\end{subfigure}
% \hfill
\begin{subfigure}{0.33\textwidth}
    \includegraphics[width=\textwidth]{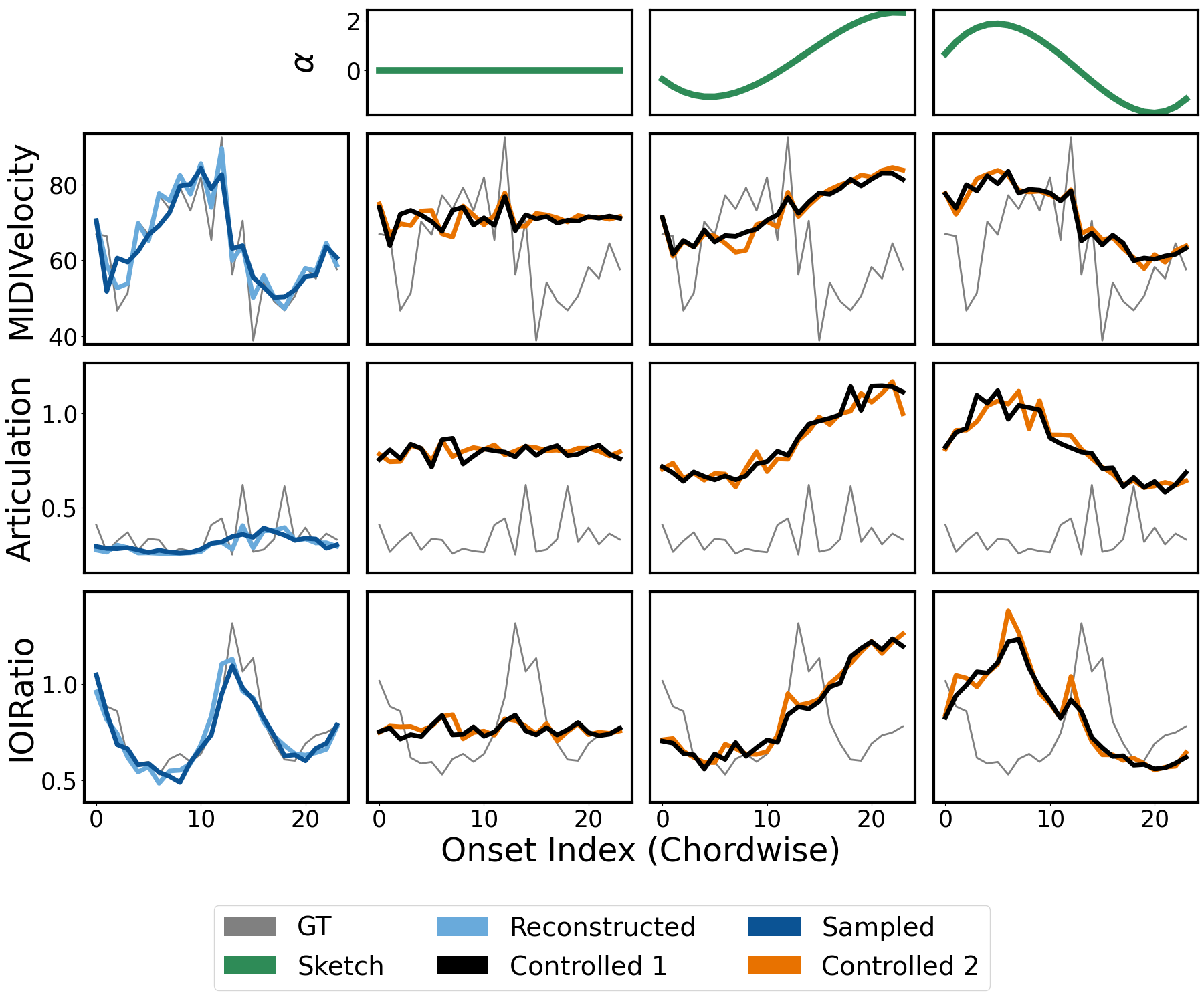}
    \caption{}
    \label{fig:sketch2}
\end{subfigure}
% \hfill
\begin{subfigure}{0.33\textwidth}
    \includegraphics[width=\textwidth]{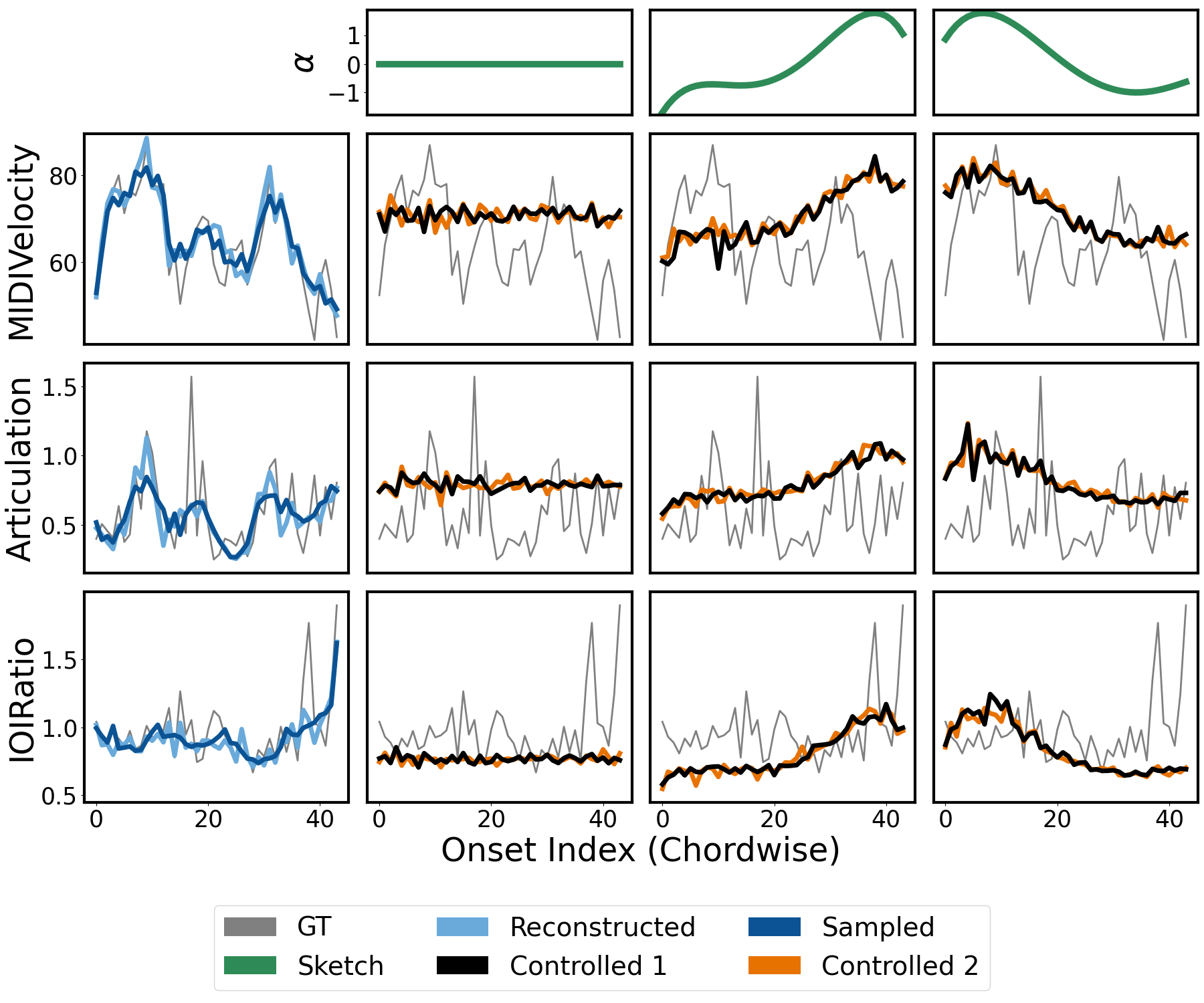}
    \caption{}
    \label{fig:sketch3}
\end{subfigure}
\caption{Qualitative samples for the proposed system. Light-blue, blue and gray lines denote the reconstructed results, sampled results from the inferred $z^{(\text{pln})}$, and their ground truths, respectively; black and orange lines denote the controlled results that are generated from different random $\tilde{z}^{(\text{str})}$; and green lines denote the "sketch" values, or $\alpha$, that are inserted to $z^{(\text{pln})}$. The samples demonstrate three excerpts that are: (a) Haydn's Keyboard Sonata, Hob. XVI:39, 3rd movement, mm. 53-56; (b) Schubert's Impromptu, Op. 90, No. 4, mm. 149-152; and (c) Balakirev's Islamey, Op. 18, mm. 29-32.}
\label{fig:sketch}
\end{figure*}

\begin{figure}
\centering
\begin{subfigure}{0.32\columnwidth}
    \includegraphics[width=\columnwidth]{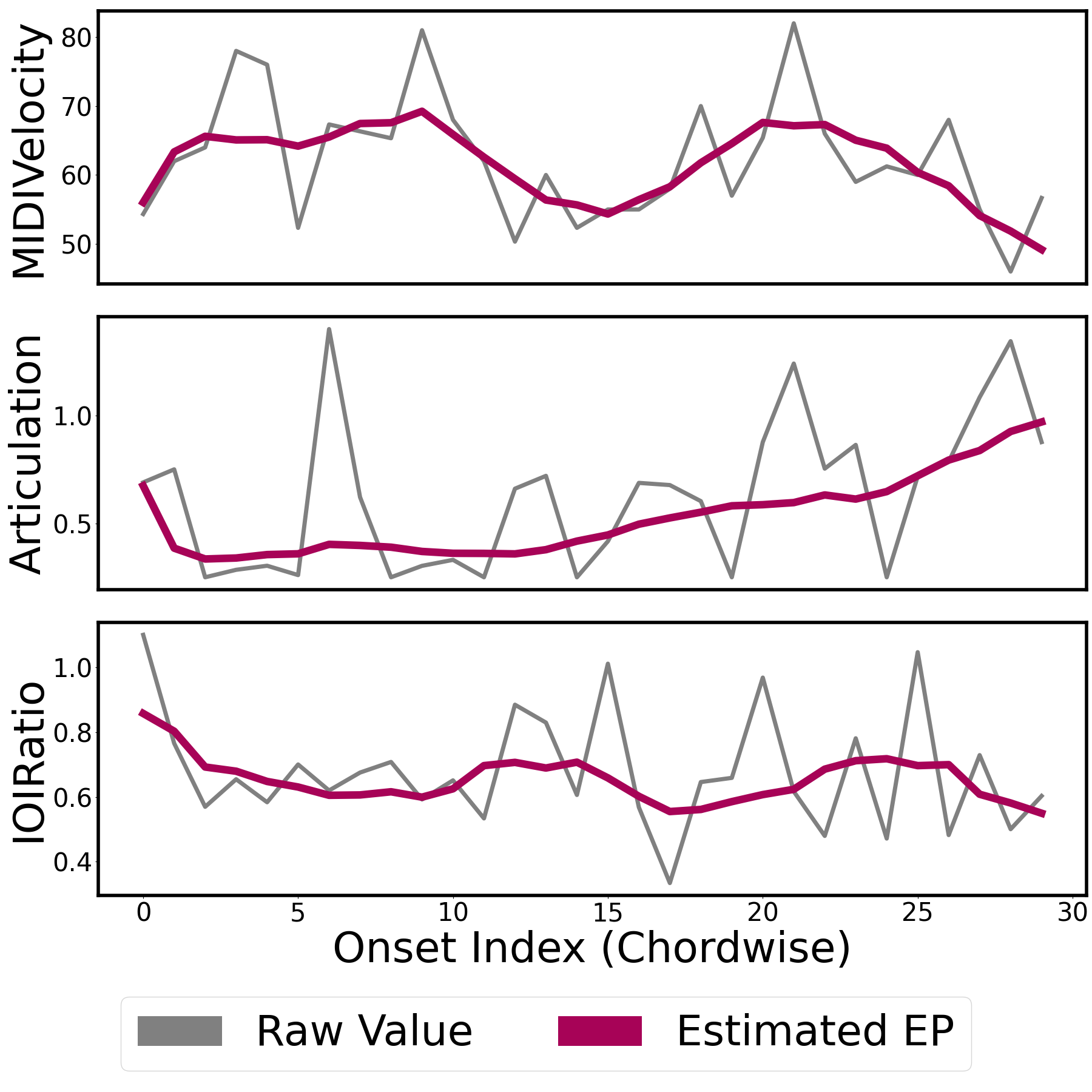}
    \caption{}
    \label{fig:ep1}
\end{subfigure}
% \hfill
\begin{subfigure}{0.32\columnwidth}
    \includegraphics[width=\columnwidth]{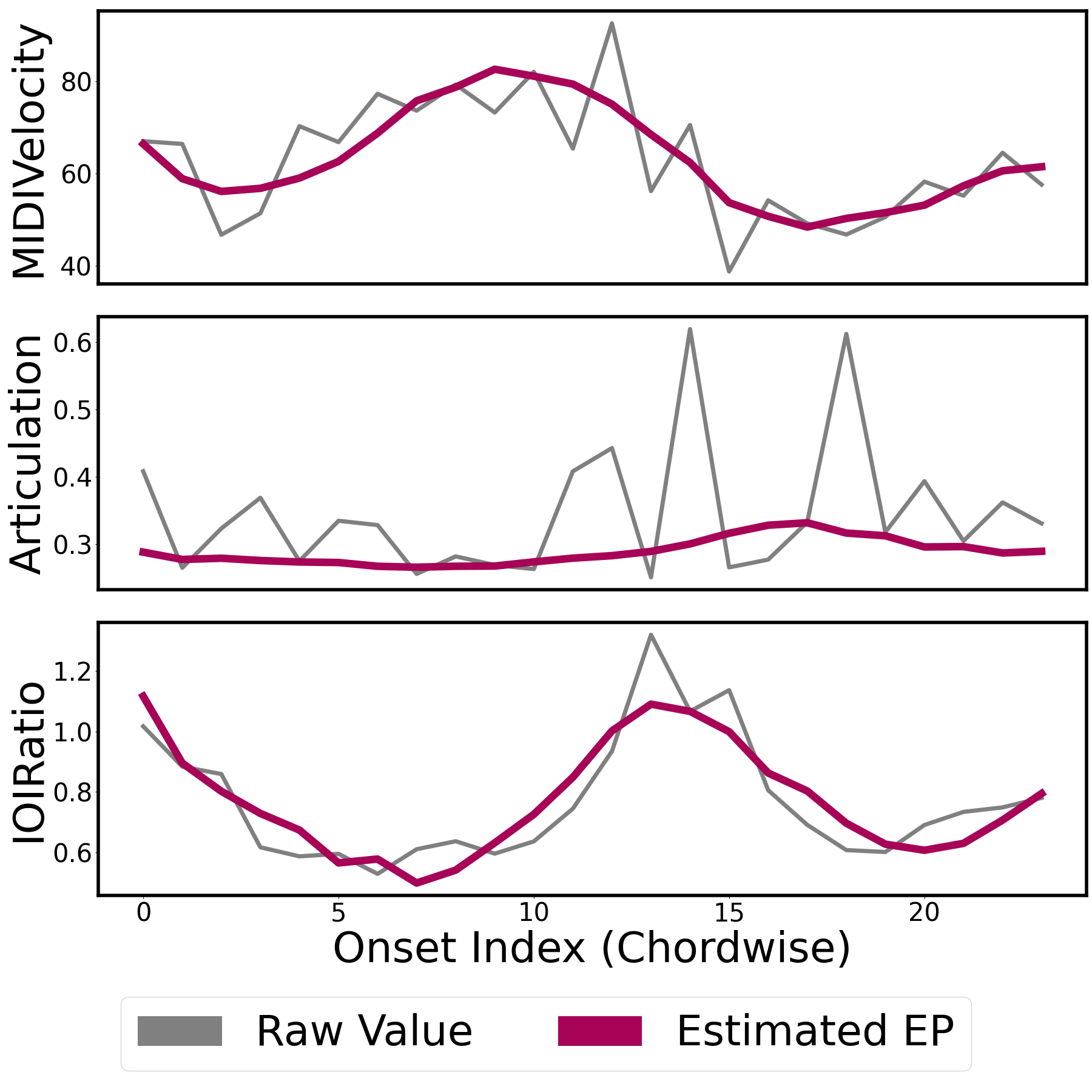}
    \caption{}
    \label{fig:ep2}
\end{subfigure}
\begin{subfigure}{0.32\columnwidth}
    \includegraphics[width=\columnwidth]{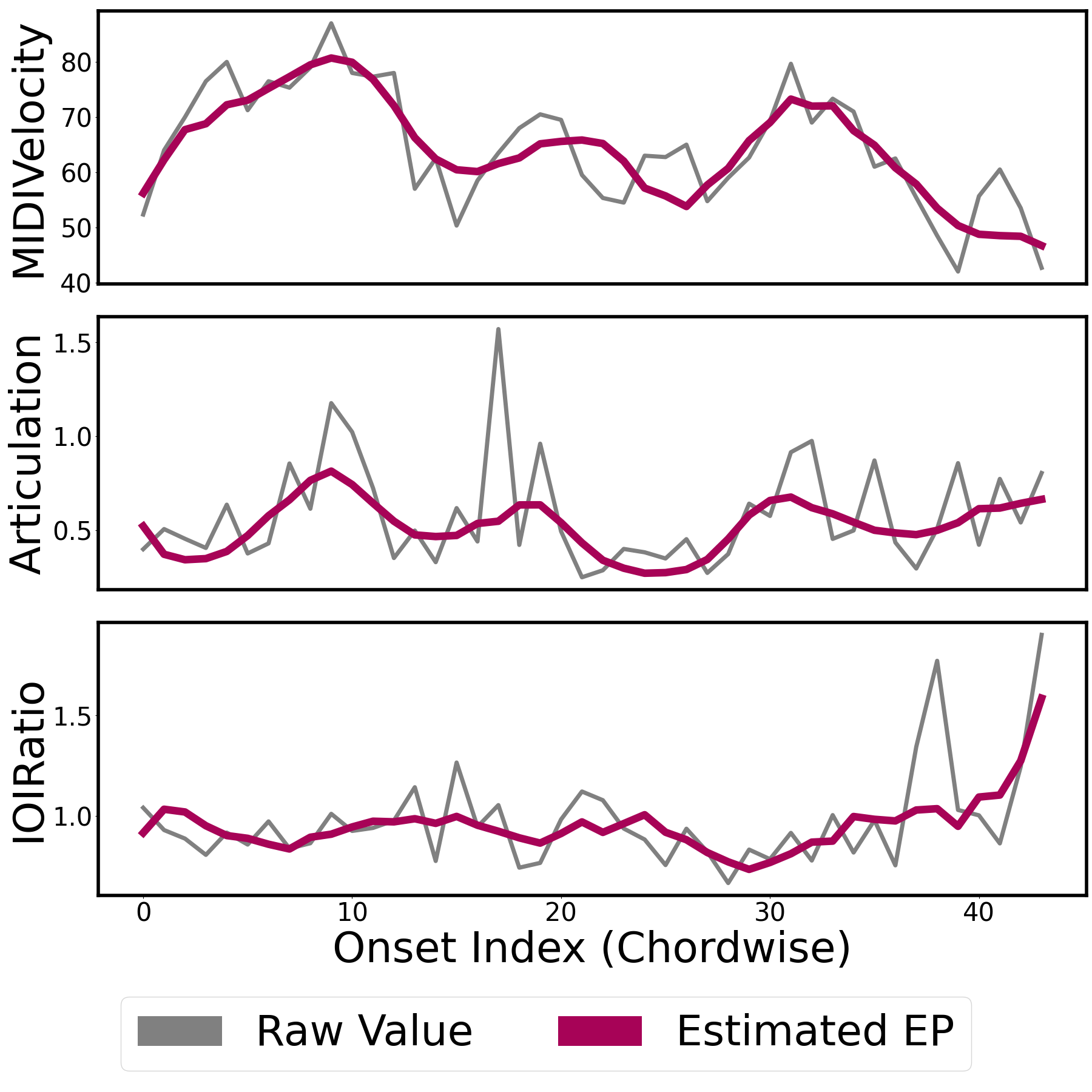}
    \caption{}
    \label{fig:ep3}
\end{subfigure}
\caption{Qualitative results for estimating the explicit planning from raw piano performances. Pink and gray lines denote the estimated contours and raw performance parameters, respectively. The results in (a), (b), and (c) are from the same excerpts for (a), (b), and (c) in \figref{fig:sketch}, respectively.}
\label{fig:ep}
\end{figure}

\subsection{Overall Objective}

The overall objective of our proposed network aims to generate realistic performance features with properly disentangled representations for the intended factors:
\begin{equation}
    \mathcal{L} = \mathcal{L}_{\text{VAE}} + \lambda_{\text{pln}}\mathcal{L}_{\text{pln}} +
    \lambda_{\text{str}}\mathcal{L}_{\text{str}} + \lambda_{\text{fac}}\mathcal{L}_{\text{fac}} + \lambda_{\text{reg}}\mathcal{L}_{\text{reg}} 
\end{equation}
where $\lambda_{\text{pln}}$, $\lambda_{\text{str}}$, $\lambda_{\text{fac}}$, and $\lambda_{\text{reg}}$ are hyperparameters for balancing the importance of the loss terms.

\section{Experimental Setups}
\label{sec:format}

% To verify whether the proposed network accomplishes the objectives, we conduct both quantitative and qualitative evaluation tasks. 
% In this section, we describe experimental settings for evaluating our proposed system. 
% First, we introduce the dataset that we use for the experiment and how the system is implemented. Next, we explain the models that are compared to our system during evaluation.

\subsection{Dataset and Implementation}

We use Yamaha e-Competition Dataset \cite{jeong2019virtuosonet} and Vienna 4x22 Piano Corpus \cite{goebl2001melody}. From these datasets, we collect 356 performances of 34 pieces by Frédéric Chopin, which have been representative research subjects for analyzing the Western musical expression \cite{repp1999microcosm,grachten2012linear,chacon2017evaluation,shi2021computational}. We use 30 pieces (108,738 batches) for training and the rest for testing. To verify the generality of model performances, we also collect the external dataset from ASAP dataset \cite{foscarin2020asap}. We use 116 performances for 23 pieces by 10 composers who represent various eras of Western music. For subjective evaluation, we collect 42 songs of non-Classical songs from online source\footnote{http://www.ambrosepianotabs.com/page/library} which are less constrained to written expression than most Classical excerpts.
% We compose each batch by slicing an entire piece into excerpts that contain maximum 16 chords where 8 chords overlap. 

We basically follow Jeong \emph{et al.} \cite{jeong2019virtuosonet} to compute the input features from the aligned pairs of performance and score data. We set MIDI velocities and Beat Per Minute (BPM) of all notes in the score data to be 64 and 120, respectively. We also remove any grace notes for simplicity and manually correct any errors. The performance features are further normalized into a range from -1 to 1 for training. We use an ADAM optimizer \cite{kingma2014adam} with an initial learning rate of 1e-5, which is reduced by 5\% every epoch during backpropagation. We empirically set $\lambda_{\text{pln}}$, $\lambda_{\text{str}}$, $\lambda_{\text{fac}}$, and $\lambda_{\text{reg}}$ to be 1000, 100, 1, 10, respectively. We set a degree of the polynomial function computing $I^{(\text{pln})}$ as 4 through an ablation study described in the supplementary material.

\subsection{Comparative Methods}

To the best of our knowledge, there is no existing method that does not intentionally follow the written guidelines in the musical score. Therefore, we use variants of our proposed network as comparing methods that differ in model architecture: {\bf Notewise} denotes the proposed model without the hierarchical learning. {\bf CVAE} denotes a variant of Notewise where $z^{(\text{pln})}$ is substituted with the supervisory signal $I^{(\text{pln})}$. We also conduct an ablation study that investigates necessity of the four loss terms.
% It is similar to Maezawa \emph{et al.}, where the model uses one latent variable with given musical conditions including dynamics and tempo \cite{maezawa2019rendering}. 

\section{Evaluation}

% We evaluate the proposed network in terms of four criteria: the generation quality, disentanglement of the latent representations, ability to control the expressive attributes, and subjective quality from human listeners.
We evaluate the proposed network in terms of both objective and subjective criteria.

\subsection{Generation Quality} 
We compute Pearson's correlation coefficients between the reconstructed or generated samples and human piano performances \cite{flossman2013expressive,chacon2017evaluation,jeong2019graph,maezawa2019rendering}. We first measure the reconstruction quality of the test samples ("$\text{R}_{\text{recon}}$"). Then, we evaluate the samples generated from $\tilde{z}^{(\text{str})}\sim p_{\theta}(z^{(\text{str})})$ and either of : 1) $z^{(\text{pln})}\sim q_{\phi}(z^{(\text{pln})}|x)$ ("$\text{R}_{x|\text{pln}}$") and 2) $z^{(\text{pln})}_{0}\sim q_{\phi}(z^{(\text{pln})}_{0}|x_{0})$ ("$\text{R}_{x|\text{pln}_{0}}$"), where $x_{0}$ is a zero matrix.

The results are shown in \tabref{tab:quan1}. Notewise shows the best scores in both datasets, and our method outperforms CVAE in $\text{R}_{\text{recon}}$. It indicates that our proposed architecture where a latent representation is used instead of a direct condition is generally good at reconstructing the human data. When using the randomly sampled $\tilde{z}^{(\text{str})}$, our method and the model without $\mathcal{L}_{\text{reg}}$ show stable scores compared to other baseline models. The model without $\mathcal{L}_{\text{reg}}$ also shows the highest scores in $\text{R}_{x|\text{pln}}$ for both datasets. It indicates that $\mathcal{L}_{\text{reg}}$ may contribute the least to generation power among other loss terms. CVAE and the model only with $\mathcal{L}^{(\text{pln})}$ also show high scores in $\text{R}_{x|\text{pln}_{0}}$. This may be due to the posterior collapse that makes the decoder depends mostly on the score condition \cite{higgins2017betavae}, which is demonstrated in the supplementary material.

\begin{table}[t]
\centering
\setlength\tabcolsep{4.5pt}
\small
{
\begin{tabular}{l|cccccc}
\hline
 \multicolumn{1}{l|}{Dataset} & \multicolumn{3}{c}{Internal} & \multicolumn{3}{c}{External} \\ \hline
 \multicolumn{1}{l|}{Metric} & $\text{R}_{\text{recon}}$ & $\text{R}_{x|\text{pln}}$ & $\text{R}_{x|\text{pln}_{0}}$ & $\text{R}_{\text{recon}}$ & $\text{R}_{x|\text{pln}}$ & $\text{R}_{x|\text{pln}_{0}}$ \\ \hline
 \multicolumn{1}{l|}{Notewise} & \textbf{0.870} & 0.392 & 0.203 & \textbf{0.875} & 0.479 & 0.177 \\ 
 \multicolumn{1}{l|}{CVAE} & 0.730 & 0.338 & 0.223 & 0.741 & 0.399 & 0.216 \\
\hline
 \multicolumn{1}{l|}{$\mathcal{L}_{\text{pln}}$} & 0.627 & 0.357 & 0.229 & 0.687 & 0.414 & \textbf{0.220}  \\
 \multicolumn{1}{l|}{$\mathcal{L}_{\text{pln}}+\mathcal{L}_{\text{str}}$} & 0.770 & 0.325 & 0.181 & 0.837 & 0.398 & 0.195 \\
 \multicolumn{1}{l|}{w/o $\mathcal{L}_{\text{fac}}$} & 0.774 & 0.289 & 0.176 & 0.838 & 0.354 & 0.173 \\ 
 \multicolumn{1}{l|}{w/o $\mathcal{L}_{\text{reg}}$} & 0.737 & \textbf{0.437} & 0.224 & 0.793 & \textbf{0.502} & 0.216 \\
\hline
 \multicolumn{1}{l|}{Ours} & 0.737 & 0.427 & \textbf{0.231} & 0.789 & 0.498 & 0.203  \\
\hline
\end{tabular}}
\caption{Evaluation results for the generation quality. The higher score is the better.}
\label{tab:quan1}
\end{table}

\begin{table}[t]
\centering
\small
{
\begin{tabular}{l|cccc}
\hline
 \multicolumn{1}{l|}{Dataset} & \multicolumn{2}{c}{Internal} & \multicolumn{2}{c}{External} \\ \hline
 \multicolumn{1}{l|}{Metric} & \makecell{$\text{MSE}_{\text{p}}$} &  \makecell{$\text{MSE}_{\text{s}}$} & \makecell{$\text{MSE}_{\text{p}}$} &  \makecell{$\text{MSE}_{\text{s}}$} \\ \hline
 \multicolumn{1}{l|}{Notewise} & 0.003 & 0.006 & 0.022 & 0.028 \\ 
 \multicolumn{1}{l|}{CVAE} & 0.034 & 0.045 & 0.085 & 0.092 \\
\hline
 \multicolumn{1}{l|}{$\mathcal{L}_{\text{pln}}$} & 0.028 & 0.036 & 0.074 & 0.077 \\
 \multicolumn{1}{l|}{$\mathcal{L}_{\text{pln}}+\mathcal{L}_{\text{str}}$} & 0.012 & 0.015 & 0.022 & 0.027 \\
 \multicolumn{1}{l|}{w/o $\mathcal{L}_{\text{fac}}$} & 0.018 & 0.023 & 0.021 & 0.025 \\ 
 \multicolumn{1}{l|}{w/o $\mathcal{L}_{\text{reg}}$} & 0.002 & 0.004 & 0.014 & 0.022 \\
\hline
 \multicolumn{1}{l|}{Ours} & \textbf{0.001} & \textbf{0.002} & \textbf{0.012} & \textbf{0.020} \\
\hline
\end{tabular}}
\caption{Evaluation results for the disentanglement of the latent representations.}
\label{tab:quan2}
\end{table}

\begin{table}[t]
\centering
\setlength\tabcolsep{4.5pt}
\small
{
\begin{tabular}{l|cccccc} \hline
 \multicolumn{1}{l|}{Dataset} & \multicolumn{3}{c}{Internal} & \multicolumn{3}{c}{External} \\ \hline
 \multicolumn{1}{l|}{Metric} & C & R & L & C & R & L \\ \hline
 \multicolumn{1}{l|}{Notewise} & 0.782 & 0.916 & 0.632 & 0.775 & 0.914 & 0.656 \\ 
 \multicolumn{1}{l|}{CVAE} & 0.798 & 0.812 & 0.620 & 0.773 & 0.802 & 0.649 \\
\hline
 \multicolumn{1}{l|}{$\mathcal{L}_{\text{pln}}$} & 0.693 & 0.852 & 0.323 & 0.694 & 0.834 & 0.324 \\
 \multicolumn{1}{l|}{$\mathcal{L}_{\text{pln}}+\mathcal{L}_{\text{str}}$} & 0.633 & 0.882 & 0.253 & 0.639 & 0.865 & 0.277 \\
 \multicolumn{1}{l|}{w/o $\mathcal{L}_{\text{fac}}$} & 0.831 & 0.846 & 0.789 & 0.832 & 0.831 & 0.847 \\ 
 \multicolumn{1}{l|}{w/o $\mathcal{L}_{\text{reg}}$} & 0.804 & \textbf{0.955} & 0.653 & 0.808 & \textbf{0.946} & 0.657 \\
\hline
 \multicolumn{1}{l|}{Ours} & \textbf{0.942} & 0.953 & \textbf{0.976} & \textbf{0.944} & 0.945 & \textbf{0.977} \\
\hline
\end{tabular}}
\caption{Evaluation results for the controllability of the expressive attributes. C, R, and L denotes consistency, restrictiveness, and linearity, respectively. Each score is the average score for the expressive attributes.}
\label{tab:quan3}
\end{table}

\begin{table}[t]
\centering
\setlength\tabcolsep{3.5pt}
\small
{
\begin{tabular}{l|ccc} \hline
 \multicolumn{1}{l|}{Metric} & \multicolumn{3}{c}{Winning Rate (Human-likeness)} \\ \hline
 \multicolumn{1}{l|}{Group} & T & UT & Overall \\ \hline
 \multicolumn{1}{l|}{Notewise} & 0.317($\pm$0.223) & 0.541($\pm$0.316) & 0.493($\pm$0.309) \\ 
 \multicolumn{1}{l|}{CVAE} & \textbf{0.467($\pm$0.356)} & 0.477($\pm$0.342) & 0.475($\pm$0.338) \\ 
 \multicolumn{1}{l|}{Ours} & 0.417($\pm$0.256) & \textbf{0.555($\pm$0.256)} & \textbf{0.525($\pm$0.258)} \\ 
 
\hline
\end{tabular}}
\caption{Evaluation results for the winning rate in terms of human-likeness. T, UT, and Overall denote musically trained, untrained, and all groups, respectively.}
\label{tab:subj}
\end{table}

% \begin{table}[t]
% \centering
% \setlength\tabcolsep{3.5pt}
% \small
% {
% \begin{tabular}{l|ccc} \hline
%  \multicolumn{1}{l|}{Metric} & \multicolumn{3}{c}{Best-score Rate} \\ \hline
%  \multicolumn{1}{l|}{Group} & T & UT & Overall \\ \hline
%  \multicolumn{1}{l|}{Notewise} & 0.000 & \textbf{0.386} & 0.304 \\ 
%  \multicolumn{1}{l|}{CVAE} & 0.417 & 0.250 & 0.286 \\ 
%  \multicolumn{1}{l|}{Ours} & \textbf{0.583} & 0.364 & \textbf{0.411} \\ 
% \hline
% \end{tabular}}
% \caption{Evaluation results for best-score rate. Overall denotes an average of all values for trained and untrained groups.}
% \label{tab:subj}
% \end{table}

\subsection{Disentangling Latent Representations}

We verify whether the latent representations are well-disentangled by appropriate information\cite{zhu2020s3vae}. To this end, each model infers the latent representations $z^{(\text{pln})}$ and $z^{(\text{str})}$ from the test sets. Each model also randomly samples $\tilde{z}^{(\text{str})}$ and infers $z^{(\text{pln})}_{0}\sim q_{\phi}(z^{(\text{pln})}|x_{0})$. We use $z^{(\text{pln})}_{0}$ to measure the structural attribute, since $z^{(\text{pln})}_{0}$ represents a flat expression where the structural attribute can be solely exposed. Each model generates new outputs as $x^{(\text{pln})}\sim p_{\theta}(x^{(\text{pln})}|z^{(\text{pln})},\tilde{z}^{(\text{str})},y)$ and $x^{(\text{str})}\sim p_{\theta}(x^{(\text{str})}|z^{(\text{pln})}_{0},z^{(\text{str})},y)$. Then, we compute a new signal $\tilde{I}^{(\text{pln})}$ from $x^{(\text{pln})}$ using the polynomial regression. The MSE values are calculated as $\text{MSE}_\text{p}=\text{MSE}(\tilde{I}^{(\text{pln})},I^{(\text{pln})})$ and $\text{MSE}_\text{s}=\text{MSE}(x^{(\text{str})},k-I^{(\text{pln})})$.

\tabref{tab:quan2} shows that our method achieves the best scores in all metrics for both datasets. This confirms that our proposed system can learn the latent representations that reflect the intended attributes. Notewise and the model without $\mathcal{L}_{\text{reg}}$ also show the robust scores compared to other baseline models. It indicates that using the notewise modeling alone is still relevant for achieving appropriate representations. It also implies that $\mathcal{L}_{\text{reg}}$ may not contribute to the disentanglement as much as other loss terms.

\subsection{Controllability of Expressive Attributes}

% We investigate whether each model can independently control the expressive attributes. To this end, 
We sample a new input $\bar{x}$ where entries of each feature are constant across time. Then, each model infers $\bar{z}^{(\text{pln})}\sim q_{\phi}(\bar{z}^{(\text{pln})}|\bar{x})$. We control each attribute by varying dimension values of $\bar{z}^{(\text{pln})}$ following Tan \emph{et al.} \cite{tan2020music} and examine the new samples generated from $\bar{z}^{(\text{pln})}$. We leverage the existing metrics to measure the controllability of each model \cite{tan2020music}: \emph{Consistency} ("C") measures consistency across samples in terms of their controlled attributes; \emph{restrictiveness} ("R") measures how much the uncontrolled attributes maintain their flatness over time; and \emph{linearity} ("L") measures how much the controlled attributes are correlated with the corresponding latent dimensions. We average over the three expressive attributes--dynamics, articulation, and tempo--into one score for each metric.

\tabref{tab:quan3} demonstrates that our system shows the best scores in consistency and linearity in both internal and external datasets. This indicates that our proposed method can robustly control the latent representation $z^{(\text{pln})}$ in intended way. The model without $\mathcal{L}_{\text{reg}}$ outperforms our method in restrictiveness. It indicates that the uncontrolled attributes by this model are the least interfered by the controlled attribute. However, its scores on consistency and linearity are lower than ours. It confirms that $\mathcal{L}_{\text{reg}}$ promotes linear control of the target attributes.

\subsection{Subjective Evaluation}

We conduct a listening test to compare the proposed model architecture to Notewise and CVAE. We qualitatively evaluate the base quality of the samples that have flat expressions, so that quality judgments are independent of any preference of arbitrary explicit planning. We generate each sample using $z^{(\text{pln})}_{0}$. A listening test is composed of 30 trials where each participant chooses a more "human-like" sample out of the generated sample and its plain MIDI \cite{jeong2019graph}. Both samples have the same length which is a maximum of 15 seconds, rendered with TiMidity++\footnote{https://sourceforge.net/projects/timidity/} without any pedal effect. \emph{Human-likeness} denotes how similar the sample is to an actual piano performance that commonly appears in popular music. A total of 28 participants are involved, and 6 participants are professionally trained in music.

The results are demonstrated in \tabref{tab:subj}. We measure a \emph{winning rate}, a rate of winning over the plain MIDI, and a \emph{top-ranking rate}, a rate of being the highest rank among the three models in terms of winning rate. These metrics are further explained in the supplementary material. The results show that musically \emph{trained} ("T") and \emph{untrained} ("UT") groups show the different tendency of each other: in the trained group, CVAE shows the best winning rate, and our method gets the best top-ranking rate; in the untrained group, our method shows the highest winning rate, whereas Notewise is top-ranked most frequently. We note that our system reveals smaller variances than those of CVAE and Notewise of the musically trained and untrained groups in the winning rate, respectively. Moreover, our system receives the highest overall scores for both metrics. It indicates that our system can be stably perceived more human-like than the plain MIDI compared to other baseline models.

\begin{figure}
\centering
\includegraphics[width=.96\columnwidth]{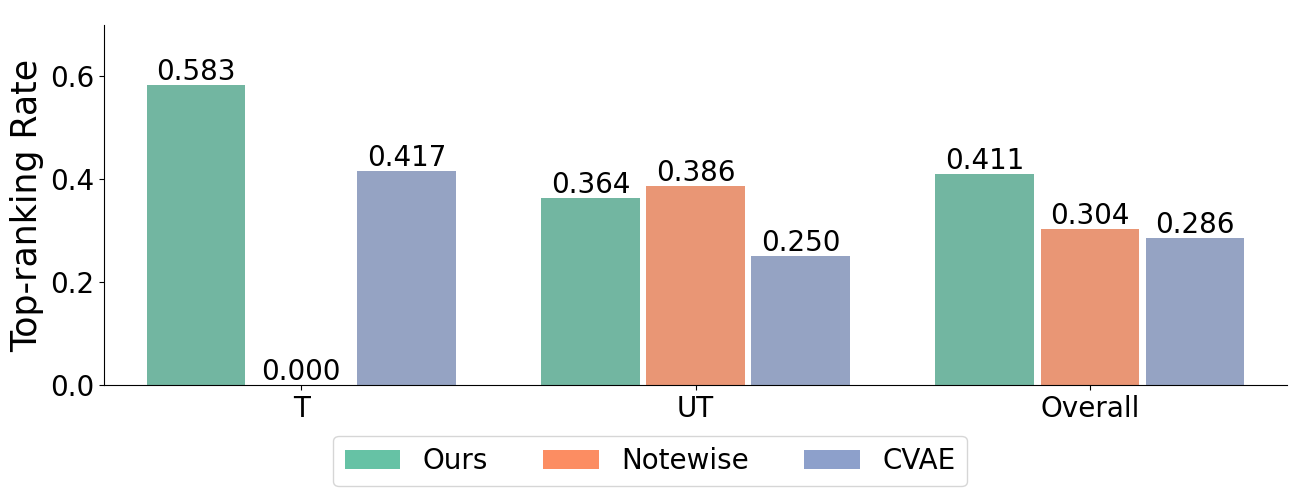}
\caption{Evaluation results for the top-ranking rate. T, UT, and Overall denote musically trained, untrained, and all groups, respectively.} % overall이 평균이 아니기 때문에 고쳐야 함!
\label{fig:best}
\end{figure}

\subsection{Qualitative Examples}  

Our system can render new piano performances from the scratch given a musical score. It can directly generate expressive parameters from the randomly sampled $\tilde{z}^{(\text{pln})}\sim p_{\theta}(z^{(\text{pln})})$ and $\tilde{z}^{(\text{str})}\sim p_{\theta}(z^{(\text{str})})$. We note that $\tilde{z}^{(\text{pln})}$ does not have temporal dependency: each $\tilde{z}^{(\text{pln})}_{c}$ is sampled independently of $\tilde{z}^{(\text{pln})}_{c-1}$. Hence, we need to insert specific values $\{\alpha^{(c)}\}_{c=1}^C$, which we call as "smooth sketches", into the target dimensions of $z^{(\text{pln})}$ if any temporal dependency of explicit planning is necessary. \figref{fig:sketch} shows that the controlled parameters are greatly correlated with $\alpha$, while their local characteristics follow those of the ground truth. In addition, the black and orange lines together demonstrate granular variety in the parameters induced by different $\tilde{z}^{(\text{str})}$ for the same musical structure. Moreover, \figref{fig:ep} shows that our system can estimate explicit planning from arbitrary human performances, indicating that our system can derive relevant information on explicit planning from the unseen data.

\section{Conclusion}

We propose a system that can render expressive piano performance with flexible control of musical expression. We attempt to achieve representations for the explicit planning and structural attribute through self-supervised learning objectives. We also leverage the two-step modeling of two hierarchical units for an intuitive generation. Experimental results confirm that our system shows stable generation quality, disentangles the target representations, and controls all expressive attributes independently of each other. Future work can be improving our system using a larger dataset for various genres and composers. We can also further compare our system with recent piano-rendering models \cite{jeong2019virtuosonet} to investigate any connections between a performer's explicit planning and a composer's intent.
% new supervisory signal $I_{\text{p}}$ with respect to whether it can be utilized to identify a performing style of music performance.
% The limitation of this study lies in using a narrow dataset for training the system. 

\section{Acknowledgments} 

We deeply appreciate Dasaem Jeong, Taegyun Kwon, and Juhan Nam for giving technical support to initiate this research. We also especially appreciate Hyeong-Seok Choi for providing critical feedback on the model architecture and evaluation. We greatly appreciate You Jin Choi and all of my colleagues who gave great help with respect to the listening test.

% For bibtex users:
% \bibliography{ISMIRtemplate}
\bibliography{ref}

\begin{thebibliography}{10}
\providecommand{\url}[1]{#1}
\csname url@samestyle\endcsname
\providecommand{\newblock}{\relax}
\providecommand{\bibinfo}[2]{#2}
\providecommand{\BIBentrySTDinterwordspacing}{\spaceskip=0pt\relax}
\providecommand{\BIBentryALTinterwordstretchfactor}{4}
\providecommand{\BIBentryALTinterwordspacing}{\spaceskip=\fontdimen2\font plus
\BIBentryALTinterwordstretchfactor\fontdimen3\font minus
  \fontdimen4\font\relax}
\providecommand{\BIBforeignlanguage}[2]{{%
\expandafter\ifx\csname l@#1\endcsname\relax
\typeout{** WARNING: IEEEtran.bst: No hyphenation pattern has been}%
\typeout{** loaded for the language `#1'. Using the pattern for}%
\typeout{** the default language instead.}%
\else
\language=\csname l@#1\endcsname
\fi
#2}}
\providecommand{\BIBdecl}{\relax}
\BIBdecl

\bibitem{widmer2004computational}
G.~Widmer and W.~Goebl, ``Computational models of expressive music performance:
  The state of the art,'' \emph{Journal of New Music Research}, vol.~33, no.~3,
  pp. 203--216, 2004.

\bibitem{chacon2018computational}
C.~E. Cancino-Chacón, M.~Grachten, W.~Goebl, and G.~Widmer, ``Computational
  models of expressive music performance: A comprehensive and critical
  review,'' \emph{Frontiers in Digital Humanities}, vol.~5, no.~25, pp. 1--23,
  2018.

\bibitem{widmer2009yqx}
G.~Widmer, S.~Flossmann, and M.~Grachten, ``{YQX} plays {C}hopin,'' \emph{AI
  Magazine}, vol.~30, no.~3, pp. 35--48, 2009.

\bibitem{kim2013statistical}
T.~H. Kim, S.~Fukayama, T.~Nishimoto, and S.~Sagayama, ``Statistical approach
  to automatic expressive rendition of polyphonic piano music,'' in \emph{Guide
  to Computing for Expressive Music Performance}.\hskip 1em plus 0.5em minus
  0.4em\relax Springer, 2013, pp. 145--179.

\bibitem{chacon2015evaluation}
C.~E. Cancino-Chacón and M.~Grachten, ``An evaluation of score descriptors
  combined with non-linear models of expressive dynamics in music,'' in
  \emph{Proceedings of the International Conference on Discovery Science},
  2015.

\bibitem{chacon2017evaluation}
C.~E. Cancino-Chacón, T.~Gadermaier, G.~Widmer, and M.~Grachten, ``An
  evaluation of linear and non-linear models of expressive dynamics in
  classical piano and symphonic music,'' \emph{Machine Learning}, vol. 106,
  no.~6, pp. 887--909, 2017.

\bibitem{maezawa2018deep}
A.~Maezawa, ``Deep piano performance rendering with conditional {VAE},'' in
  \emph{Late-Breaking Demo, the 19th International Society for Music
  Information Retrieval Conference}, 2018.

\bibitem{jeong2019virtuosonet}
D.~Jeong, T.~Kwon, Y.~Kim, K.~Lee, and J.~Nam, ``Virtuoso{N}et: A hierarchical
  {RNN}-based system for modeling expressive piano performance,'' in
  \emph{Proceedings of the 20th International Society for Music Information
  Retrieval}, 2019.

\bibitem{jeong2019graph}
D.~Jeong, T.~Kwon, Y.~Kim, and J.~Nam, ``Graph neural network for music score
  data and modeling expressive piano performance,'' in \emph{Proceedings of the
  36th International Conference on Machine Learning}, 2019.

\bibitem{bengio2013representation}
Y.~Bengio, A.~Courville, and P.~Vincent, ``Representation learning: A review
  and new perspectives,'' \emph{IEEE Transactions on Pattern Analysis and
  Machine Intelligence}, vol.~35, no.~8, 2013.

\bibitem{maezawa2019rendering}
A.~Maezawa, K.~Yamamoto, and T.~Fujishima, ``Rendering music performance with
  interpretation variations using conditional variational {RNN},'' in
  \emph{Proceedings of the 20th International Society for Music Information
  Retrieval Conference}, 2019.

\bibitem{tan2020generating}
H.~H. Tan, Y.-J. Luo, and D.~Herremans, ``Generative modeling for controllable
  audio synthesis of expressive piano performance,'' in \emph{Proceedings of
  the 37th International Conference on Machine Learning}, 2020.

\bibitem{bresin2000emotional}
R.~Bresin and A.~Friberg, ``Emotional coloring of computer-controlled music
  performances,'' \emph{Computer Music Journal}, vol.~24, no.~4, 2000.

\bibitem{livingstone2010changing}
S.~R. Livingstone, R.~Muhlberger, A.~R. Brown, and W.~F. Thompson, ``Changing
  musical emotion: A computational rule system for modifying score and
  performance,'' \emph{Computer Music Journal}, vol.~34, no.~1, 2010.

\bibitem{bernays2014investigating}
M.~Bernays and C.~Traube, ``Investigating pianists’ individuality in the
  performance of five timbral nuances through patterns of articulation, touch,
  dynamics, and pedaling,'' \emph{Frontiers in Psychology}, vol.~5, no. 157,
  pp. 1--19, 2014.

\bibitem{oore2020thistime}
S.~Oore, I.~Simon, S.~Dieleman, D.~Eck, and K.~Simonyan, ``This time with
  feeling: Learning expressive musical performance,'' \emph{Neural Computing
  and Applications}, vol.~32, pp. 955--967, 2020.

\bibitem{bhatara2011perception}
A.~Bhatara, A.~K. Tirovolas, L.~M. Duan, B.~Levy, and D.~J. Levitin,
  ``Perception of emotional expression in musical performance,'' \emph{Journal
  of Experimental Psychology: Human Perception and Performance}, vol.~37,
  no.~3, pp. 921--934, 2011.

\bibitem{friberg2006overview}
A.~Friberg, R.~Bresin, and J.~Sundberg, ``Overview of the {KTH} rule system for
  musical performance,'' \emph{Advances in Cognitive Psychology}, vol.~2, no.
  2-3, pp. 145--161, 2006.

\bibitem{flossman2013expressive}
S.~Flossmann, M.~Grachten, and G.~Widmer, ``Expressive performance rendering
  with probabilistic models,'' in \emph{Guide to Computing for Expressive Music
  Performance}.\hskip 1em plus 0.5em minus 0.4em\relax Springer, 2013, pp.
  75--98.

\bibitem{woody1999relationship}
R.~H. Woody, ``The relationship between explicit planning and expressive
  performance of dynamic variations in an aural modeling task,'' \emph{Journal
  of Research in Music Education}, vol.~47, no.~4, pp. 331--342, 1999.

\bibitem{lerch2019music}
A.~Lerch, C.~Arthur, A.~Pati, and S.~Gururani, ``Music performance analysis: A
  survey,'' in \emph{Proceedings of the 20st International Society for Music
  Information Retrieval Conference}, 2019.

\bibitem{repp1999microcosm}
B.~H. Repp, ``A microcosm of musical expression: {II}. quantitative analysis of
  pianists’ dynamics in the initial measures of {C}hopin’s {E}tude in {E}
  major,'' \emph{The Journal of the Acoustical Society of America}, vol. 105,
  no.~3, pp. 1972--1988, 1999.

\bibitem{honing2001from}
H.~Honing, ``From time to time: The representation of timing and tempo,''
  \emph{Computer Music Journal}, vol.~25, no.~3, 2001.

\bibitem{zhu2020s3vae}
Y.~Zhu, M.~R. Min, A.~Kadav, and H.~P. Graf, ``{S3VAE}: Self-supervised
  sequential {VAE} for representation disentanglement and data generation,'' in
  \emph{Proceedings of Computer Vision and Pattern Recognition}, 2020.

\bibitem{locatello2019challenging}
F.~Locatello, S.~Bauer, M.~Lucic, G.~Rätsch, S.~Gelly, B.~Schölkopf, and
  O.~Bachem, ``Challenging common assumptions in the unsupervised learning of
  disentangled representations,'' in \emph{Proceedings of the 36th
  International Conference on Machine Learning}, 2019.

\bibitem{hendrycks2019using}
D.~Hendrycks, M.~Mazeika, S.~Kadavath, and D.~Song, ``Using self-supervised
  learning can improve model robustness and uncertainty,'' in \emph{Proceedings
  of the 33rd Conference on Neural Information Processing Systems}, 2019.

\bibitem{pati2020attribute}
A.~Pati and A.~Lerch, ``Attribute-based regularization of latent spaces for
  variational auto-encoders,'' in \emph{Neural Computing and Applications},
  2020.

\bibitem{tan2020music}
H.~H. Tan and D.~Herremans, ``Music {F}ader{N}ets: Controllable music
  generation based on high-level features via low-level feature modeling,'' in
  \emph{Proceedings of the 21st International Society for Music Information
  Retrieval Conference}, 2020.

\bibitem{jeong2019score}
D.~Jeong, T.~Kwon, Y.~Kim, and J.~Nam, ``Score and performance features for
  rendering expressive music performances,'' in \emph{Proceedings of the Music
  Encoding Conference}, 2019.

\bibitem{roberts2017hierarchical}
A.~Roberts, J.~Engel, and D.~Eck, ``Hierarchical variational autoencoders for
  music,'' in \emph{Proceedings of the 31st Conference on Neural Information
  Processing Systems}, 2017.

\bibitem{dong2018musegan}
H.-W. Dong, W.-Y. Hsiao, L.-C. Yang, and Y.-H. Yang, ``Muse{GAN}: Multi-track
  sequential generative adversarial networks for symbolic music generation and
  accompaniment,'' in \emph{Proceedings of the 32nd AAAI Conference on
  Artificial Intelligence}, 2018.

\bibitem{wu2021musemorphose}
S.-L. Wu and Y.-H. Yang, ``Muse{M}orphose: Full-song and fine-grained music
  style transfer with just one {T}ransformer {VAE},'' \emph{arXiv preprint
  arXiv:2105.04090}, 2021.

\bibitem{wang2020pianotree}
Z.~Wang, Y.~Zhang, Y.~Zhang, J.~Jiang, R.~Yang, J.~Z. (Jake), and G.~Xia,
  ``Pianotree {VAE}: Structured representation learning for polyphonic music,''
  in \emph{Proceedings of the 37th International Conference on Machine
  Learning}, 2020.

\bibitem{kingma2013auto}
D.~P. Kingma and M.~Welling, ``Auto-encoding variational bayes,'' \emph{arXiv
  preprint arXiv:1312.6114}, 2013.

\bibitem{sohn2015learning}
K.~Sohn, X.~Yan, and H.~Lee, ``Learning structured output representation using
  deep conditional generative models,'' in \emph{Proceedings of the 28th
  International Conference on Neural Information Processing Systems}, 2015.

\bibitem{li2018disentangled}
Y.~Li and S.~Mandt, ``Disentangled sequential autoencoder,'' in
  \emph{Proceedings of the 35th International Conference on Machine Learning},
  2018.

\bibitem{chung2015recurrent}
J.~Chung, K.~Kastner, L.~Dinh, K.~Goel, A.~C. Courville, and Y.~Bengio, ``A
  recurrent latent variable model for sequential data,'' in \emph{Proceedings
  of the 28th International Conference on Neural Information Processing
  Systems}, 2015.

\bibitem{williams1989learning}
R.~J. Williams and D.~Zipser, ``A learning algorithm for continually running
  fully recurrent neural networks,'' \emph{Neural Computation}, vol.~1, no.~2,
  1989.

\bibitem{hu2017toward}
Z.~Hu, Z.~Yang, X.~Liang, R.~Salakhutdinov, and E.~P. Xing, ``Toward controlled
  generation of text,'' in \emph{Proceedings of the 34th International
  Conference on Machine Learning}, 2017.

\bibitem{goebl2001melody}
W.~Goebl, ``Melody lead in piano performance: Expressive device or artifact?''
  \emph{The Journal of the Acoustical Society of America}, vol. 110, no.~1,
  2001.

\bibitem{grachten2012linear}
M.~Grachten and G.~Widmer, ``Linear basis models for prediction and analysis of
  musical expression,'' \emph{Journal of New Music Research}, vol.~41, no.~4,
  pp. 311--322, 2012.

\bibitem{shi2021computational}
Z.~Shi, ``Computational analysis and modeling of expressive timing in {C}hopin
  {M}azurkas,'' in \emph{Proceedings of the 22nd International Society for
  Music Information Retrieval Conference}, 2021.

\bibitem{foscarin2020asap}
F.~Foscarin, A.~McLeod, P.~Rigaux, F.~Jacquemard, and M.~Sakai, ``{ASAP}: A
  dataset of aligned scores and performances for piano transcription,'' in
  \emph{Proceedings of the 21st International Society for Music Information
  Retrieval Conference}, 2020.

\bibitem{kingma2014adam}
D.~P. Kingma and J.~Ba, ``Adam: A method for stochastic optimization,''
  \emph{arXiv preprint arXiv:1412.6980}, 2014.

\bibitem{higgins2017betavae}
I.~Higgins, L.~Matthey, A.~Pal, C.~Burgess, X.~Glorot, M.~Botvinick,
  S.~Mohamed, and A.~Lerchner, ``\textbeta-{VAE}: Learning basic visual
  concepts with a constrained variational framework,'' in \emph{Proceedings of
  the 5th International Conference on Learning Representations}, 2017.

\end{thebibliography}

% For non bibtex users:
%\begin{thebibliography}{citations}
% \bibitem{Author:17}
% E.~Author and B.~Authour, ``The title of the conference paper,'' in {\em Proc.
% of the Int. Society for Music Information Retrieval Conf.}, (Suzhou, China),
% pp.~111--117, 2017.
%
% \bibitem{Someone:10}
% A.~Someone, B.~Someone, and C.~Someone, ``The title of the journal paper,''
%  {\em Journal of New Music Research}, vol.~A, pp.~111--222, September 2010.
%
% \bibitem{Person:20}
% O.~Person, {\em Title of the Book}.
% \newblock Montr\'{e}al, Canada: McGill-Queen's University Press, 2021.
%
% \bibitem{Person:09}
% F.~Person and S.~Person, ``Title of a chapter this book,'' in {\em A Book
% Containing Delightful Chapters} (A.~G. Editor, ed.), pp.~58--102, Tokyo,
% Japan: The Publisher, 2009.
%
%
%\end{thebibliography}

\end{document}

\begin{appendices}
%
% \maketitle
%
%
\section{Experimental Setups}

\subsection{Metrics Detail}

{\bf Controllability.} We utilize the existing metrics which are \emph{consistency}, \emph{restrictiveness}, and \emph{linearity} to investigate whether the expressive attributes can be independently controlled \cite{tan2020music}. For controlling each attribute, each model first infers $z^{(\text{pln})}$ for all test samples. Then, we compute the maximum and minimum values of the target dimension $d^{(\text{attr})}={\{d_{t}^{(\text{attr})}\}_{t=1}^{T}}$, where $T\in \{N,C\}$. Then, we set a dimension value of each timestep as $d^{(\text{attr})}_{t}=\text{min}(d^{(\text{attr})})+\frac{t}{T}(\text{max}(d^{(\text{attr})})-\text{min}(d^{(\text{attr})}))$. If an expressive attribute is controlled by $z^{(\text{pln})}$ in the appropriate way, the corresponding attribute should only change along with the target dimension values while the other attributes maintain their status. In the case of controlling dynamics of the samples, the metrics are computed as follows \cite{tan2020music}:
\begin{equation}
    \text{Consistency} = 1 - \frac{1}{T}\sum_{t=1}^T\underset{t}{\sigma}(v_{1...M,t})
\end{equation}
\begin{equation}
\begin{aligned}
    &\text{Restrictiveness} \\
    &= 1-\frac{1}{2M}\left(\sum_{m=1}^M\underset{m}{\sigma}(a_{m,1..T}) + \underset{m}{\sigma}(i_{m,1..T})\right)
\end{aligned}
\end{equation}
\begin{equation}
    \text{Linearity} = R^2(\mathcal{S}(p_{1...M}))
\end{equation}
where $v$, $a$, and $i$ are respectively the values of \textit{MIDIVelocity}, \textit{Articulation}, and \textit{IOIRatio} of the generated output, $\mathcal{S}$ is a linear regression model, $p_{m}=\{(d^{(\text{attr})}_{t}, v_{m,t})|t\in[1,T]\}$, and $M$ is the number of samples.

{\bf Subjective Evaluation.} We measure subjective quality of the generated samples in terms of \emph{winning rate} and \emph{top-ranking rate}. Winning rate is computed for each model as a ratio of the number of winning the plain MIDI to the total number of trials per each participant. Top-ranking rate is calculated as a ratio of the number of participants who choose the corresponding model most frequently among the three models to the total number of participants in each group. Concretely, the number of being top-ranked by each participant is counted by $1/numModel$, where $numModel$ denotes the number of models that are being top-ranked at the same time.

\subsection{Implementation Details}

We compose each batch by slicing an entire piece into short excerpts where notes for maximum 16 chords are contained and 12 chords overlap. Thus, a length of each batch varies from 16 to 114. 
The embedding sizes of the performance input $x$ and score input $y$ are 256 and 128, respectively. The sizes of $z^{(\text{pln})}$, $z^{(\text{str})}$, and hidden dimension are 12 and 64, and 256, respectively. We train all models for 100 epochs (170,000 iterations) with a batch size of 64. For quantitative evaluation, each model repeatedly generates 20 samples for the same inputs considering the randomness of each result. For subjective and qualitative evaluations, all models generate the samples using the truncation trick with a threshold of 2 \cite{brock2017large}.

\begin{table}[t]
\centering
\setlength\tabcolsep{4.5pt}
\small
{
\begin{tabular}{l|cc}
\hline
 \multicolumn{1}{l|}{Dataset} & \multicolumn{2}{c}{Internal} \\ \hline
 \multicolumn{1}{l|}{Metrics} & $\text{KLD}_{\text{p}}$ & $\text{KLD}_{\text{s}}$ \\ \hline
 \multicolumn{1}{l|}{Notewise} & 1.2601($\pm$0.3141) & 0.5294($\pm$0.0850) \\ 
 \multicolumn{1}{l|}{CVAE} & - & \textbf{0.0225($\pm$0.0052)} \\
\hline
 \multicolumn{1}{l|}{$\mathcal{L}_{\text{pln}}$} & 0.9139($\pm$0.1274) & 0.0268($\pm$0.0049) \\
 \multicolumn{1}{l|}{$\mathcal{L}_{\text{pln}}+\mathcal{L}_{\text{str}}$} & 0.9468($\pm$0.1210) & 0.6041($\pm$0.0441)  \\
 \multicolumn{1}{l|}{w/o $\mathcal{L}_{\text{fac}}$} & 1.1374($\pm$0.4421) & 0.5127($\pm$0.0395) \\ 
 \multicolumn{1}{l|}{w/o $\mathcal{L}_{\text{reg}}$} & \textbf{0.8856($\pm$0.1759)} & 0.6597($\pm$0.0545) \\
\hline
 \multicolumn{1}{l|}{Ours} & 1.4338($\pm$0.7836) & 0.6298($\pm$0.0495) \\
\hline
\hline
 \multicolumn{1}{l|}{Dataset} & \multicolumn{2}{c}{External} \\ \hline
 \multicolumn{1}{l|}{Metrics} & $\text{KLD}_{\text{p}}$ & $\text{KLD}_{\text{s}}$ \\ \hline
 \multicolumn{1}{l|}{Notewise} & 1.1671($\pm$0.4194) & 0.4559($\pm$0.1084) \\ 
 \multicolumn{1}{l|}{CVAE} & - & \textbf{0.0188($\pm$0.0055)} \\
\hline
 \multicolumn{1}{l|}{$\mathcal{L}_{\text{pln}}$} & 1.0053($\pm$0.2099) & 0.0278($\pm$0.0061) \\
 \multicolumn{1}{l|}{$\mathcal{L}_{\text{pln}}+\mathcal{L}_{\text{str}}$} & 1.0528($\pm$0.2610) & 0.6260($\pm$0.0539)  \\
 \multicolumn{1}{l|}{w/o $\mathcal{L}_{\text{fac}}$} & 1.4065($\pm$0.6481) & 0.5329($\pm$0.0474) \\ 
 \multicolumn{1}{l|}{w/o $\mathcal{L}_{\text{reg}}$} & \textbf{0.9773($\pm$0.2403)} & 0.6688($\pm$0.0602) \\
\hline
 \multicolumn{1}{l|}{Ours} & 1.7641($\pm$0.9871) & 0.6337($\pm$0.0581) \\
\hline
\end{tabular}}
\caption{Evaluation results for the KL divergence loss.}
\label{tab:quan_a1}
\end{table}

\begin{table}[t]
\centering
\setlength\tabcolsep{4pt}
\small
{
\begin{tabular}{l|cccccc}
\hline
 \multicolumn{1}{l|}{Dataset} & \multicolumn{3}{c}{Internal} & \multicolumn{3}{c}{External} \\ \hline
 \multicolumn{1}{l|}{Metrics} & $\text{R}_{\text{recon}}$ & $\text{R}_{\text{x}|\text{pln}}$ & $\text{R}_{\text{x}|\text{pln}_{0}}$ & $\text{R}_{\text{recon}}$ & $\text{R}_{\text{x}|\text{pln}}$ & $\text{R}_{\text{x}|\text{pln}_{0}}$ \\ \hline
 \multicolumn{1}{l|}{$d_{I^{(\text{pln})}}=1$} & 0.735 & 0.298 & 0.218 & 0.784 & 0.330 & \textbf{0.214} \\
 \multicolumn{1}{l|}{$d_{I^{(\text{pln})}}=2$} & 0.717 & 0.348 & 0.225 & 0.769 & 0.394 & 0.204 \\ 
 \multicolumn{1}{l|}{$d_{I^{(\text{pln})}}=4$} & \textbf{0.737} & 0.427 & \textbf{0.231} & \textbf{0.789} & 0.498 & 0.203  \\
 \multicolumn{1}{l|}{$d_{I^{(\text{pln})}}=8$} & 0.719 & \textbf{0.546} & 0.197 & 0.786 & \textbf{0.650} & 0.211 \\
\hline
\end{tabular}}
\caption{Evaluation results for the generation quality. The higher score is the better.}
\label{tab:quan_a2}
\end{table}

\begin{table}[t]
\centering
\small
{
\begin{tabular}{l|cccc}
\hline
 \multicolumn{1}{l|}{Dataset} & \multicolumn{2}{c}{Internal} & \multicolumn{2}{c}{External} \\ \hline
 \multicolumn{1}{l|}{Metrics} & \makecell{$\text{MSE}_{\text{p}}$} &  \makecell{$\text{MSE}_{\text{s}}$} & \makecell{$\text{MSE}_{\text{p}}$} &  \makecell{$\text{MSE}_{\text{s}}$} \\ \hline
 \multicolumn{1}{l|}{$d_{I^{(\text{pln})}}=1$} & \textbf{0.0013} & 0.0150 & \textbf{0.0021} & 0.0229 \\ 
 \multicolumn{1}{l|}{$d_{I^{(\text{pln})}}=2$} & 0.0024 & 0.0149 & 0.0028 & 0.0243 \\
 \multicolumn{1}{l|}{$d_{I^{(\text{pln})}}=4$} & \textbf{0.0013} & \textbf{0.0115} & \textbf{0.0021} & 0.0196 \\
 \multicolumn{1}{l|}{$d_{I^{(\text{pln})}}=8$} & 0.0017 & 0.0127 & 0.0022 & \textbf{0.0172} \\
\hline
\end{tabular}}
\caption{Evaluation results for the disentanglement of the latent representations.}
\label{tab:quan_a3}
\end{table}

\section{Quantitative Results}

\subsection{KL Divergence}

\tabref{tab:quan_a1} shows the results for the KL divergence of $z^{(\text{pln})}$ and $z^{(\text{str})}$ which we denote as "$\text{KLD}_{\text{p}}$" and "$\text{KLD}_{\text{s}}$", respectively. It shows that our method reveals the highest KL divergence of both latent variables in both datasets. In contrast, the model without $\mathcal{L}_{\text{reg}}$ and CVAE shows the lowest values for $\text{KLD}_{\text{p}}$ and $\text{KLD}_{\text{s}}$, respectively. In particular, CVAE and the model only with $\mathcal{L}_{\text{pln}}$ show abrupt decreases in $\text{KLD}_{\text{s}}$ compared to other models. It shows that these models have extremely small regularization power: it may have led to the posterior collapse of any information that $z^{(\text{str})}$ should carry and allowed the decoder to become mostly dependent on the deterministic condition of the score features \cite{higgins2017betavae}.

\subsection{Ablation Study}

We also conduct an ablation study for the degree of the polynomial function to compute $I^{(\text{pln})}$. We investigate the cases where the degree $d_{I^{(\text{pln})}}$ is 1, 2, 4, or 8. \tabref{tab:quan_a2} and \tabref{tab:quan_a3} show the results for the metrics of the generation quality and disentanglement of the representations, respectively. In the generation quality, $d_{I^{(\text{pln})}}=4$ receives the best scores for the three metrics out of the six, compared to other models. In particular, $d_{I^{(\text{pln})}}=4$ shows the highest reconstruction scores in both datasets, whereas $d_{I^{(\text{pln})}}=8$ shows the best scores for $R_{x|\text{pln}}$ in both datasets. $d_{I^{(\text{pln})}}=4$ also shows the best score for $R_{x|\text{pln}_{0}}$ in the internal dataset. However, $d_{I^{(\text{pln})}}=1$ is the highest for $R_{x|\text{pln}_{0}}$ in the external dataset instead of $d_{I^{(\text{pln})}}=8$. In the disentanglement metrics, nonetheless, our method with $d_{I^{(\text{pln})}}=4$ shows the best scores for most metrics. The model with $d_{I^{(\text{pln})}}=1$ shows the best scores for $\text{MSE}_{\text{p}}$ in both datasets but relatively low scores for $\text{MSE}_{\text{s}}$. According to these results, we determine the degree of the polynomial function for $I^{(\text{pln})}$ to be 4 in this study.

\section{More Qualitative Samples}

% \subsection{Interpolation}

% We illustrate several sets of samples generated by interpolating between a pair of $z_{\text{P}}$ inferred from a pair of two performance samples. We first sample three pairs two piano performances by increasing or decreasing the original values of dynamics, articulation, or tempo by $\pm30\%$. We denote the resulted pairs as: $\{\text{loud}, \text{quiet}\}$, $\{\text{staccato}, \text{legato}\}$, or $\{\text{fast}, \text{slow}\}$, respectively. Next, $\{z_{\text{P}_{a}}, z_{\text{P}_{b}}\}$ is inferred from two samples in a pair, where $a$ and $b$ are the corresponding samples. We interpolate between $\{z_{\text{P}_{a}}, z_{\text{P}_{b}}\}$ and produce its mixture as $z_{\text{P}}^{a,b}=z_{\text{P}_{a}}\times 0.5 + z_{\text{P}_{b}}\times 0.5$. Figure \ref{fig:intp} show the results generated from $\{z_{\text{P}_{a}}, z_{\text{P}}^{a,b}, z_{\text{P}_{b}}\}$ of each condition pair. The samples are based on the same random $z_{\text{S}}$. The results illustrate that only the target attribute changes along the direction of interpolation, while other attributes maintain their original status. It demonstrates that $z_{\text{P}}$ within our model carries appropriate, independent information with respect to the explicit planning for the expressive attribute.

% \subsection{More Qualitative Samples}

Demo and more qualitative samples are introduced in the online page \footnote{\url{https://free-pig-6c6.notion.site/DEMO-c20a1fea7a0844468a05b971c3b9ef3c}}

\end{appendices}

% For bibtex users:
\bibliography{ref}